\documentclass{desyproc}

\usepackage{mathrsfs}
\usepackage{hyperref}

\begin{document}
\title{Inferences on the Higgs Boson and Axion Masses\\ through a Maximum Entropy Principle}

\author{{\slshape Alexandre Alves$^{1}$, Alex G. Dias$^{2}$, Roberto da Silva$^{3}$}\\[1ex]
$^1$Federal University of S\~ao Paulo, Physics Department, Diadema, Brazil\\
$^2$Federal University of ABC, Center of Natural and Human Sciences, Santo Andr\'e, Brazil\\
$^3$Federal University of Rio Grande do Sul, Physics Institute, Porto Alegre, Brazil}

\contribID{familyname\_firstname}

\confID{13889}  
\desyproc{DESY-PROC-2017-XX}
\acronym{Patras 2017} 
\doi  

\maketitle

\begin{abstract}
The Maximum Entropy Principle (MEP) is a method that can be used to infer the 
value of an unknown quantity in a set of probability functions. 
In this work we review two applications of MEP: one giving a precise inference of 
the Higgs boson mass value; and the other one allowing to infer the mass of 
the axion. In particular, for the axion we assume that it has a decay channel 
into pairs of neutrinos, in addition to the decay into two photons. 
The Shannon entropy associated to an initial ensemble of axions decaying into 
photons and neutrinos is then built for maximization. 
\end{abstract}

\section*{The Method of Maximum Entropy Principle}

Among the biggest challenges of physics is to find an explanation for the values of 
the masses of the elementary particles. With the discovery of the Higgs boson at the LHC, 
and the measurements of its properties, we have now the experimental evidence 
that the mechanism of 
spontaneous symmetry breaking is effectively behind the mass generation for the 
elementary particles, as implemented in the Standard Model (SM). 
Although this represented an invaluable 
advance for understanding of subatomic physics, the spontaneous symmetry breaking 
mechanism in the SM does not determine the value of the masses of the elementary 
particles, except for the photon which is correctly left massless and all the neutrinos 
which are incorrectly left massless. It means that some still missing mechanism must 
supplement, or totally replace, the spontaneous symmetry breaking in order to allow 
a full determination of the masses of  the  elementary particles.

In particular, for the Higgs boson some ideas have arised relating its mass 
to the maximum of a  probability distribution built from the branching 
ratios~\cite{dEnterria:2012eip},~\cite{Alves:2014ksa}. 
As observed in Ref.~\cite{dEnterria:2012eip} a probability 
distribution function constructed multiplying the main branching ratios for 
the SM Higgs boson decay has a peak near to the measured value 
$M_H\approx 125$ GeV. This could be connected to some sort of entropy so that 
such a mass value maximizes, simultaneously, the decay 
probabilities into all SM particles. In the  
work of Ref.~\cite{Alves:2014ksa} we presented a development based on the Shannon 
entropy  built with the Higgs boson branching ratios, showing that 
the value of the Higgs boson mass follows from a Maximum Entropy Principle (MEP). 

In the following we present a short review of our MEP method to infer the mass of the 
SM Higgs boson. After that we also review an application of MEP for 
inferring the mass of the axion assuming a specific low energy 
effective Lagrangian~\cite{Alves:2017ljt}. 

\subsection*{Inferring the Higgs boson mass}

\begin{figure}
  \center \includegraphics[scale=0.36]{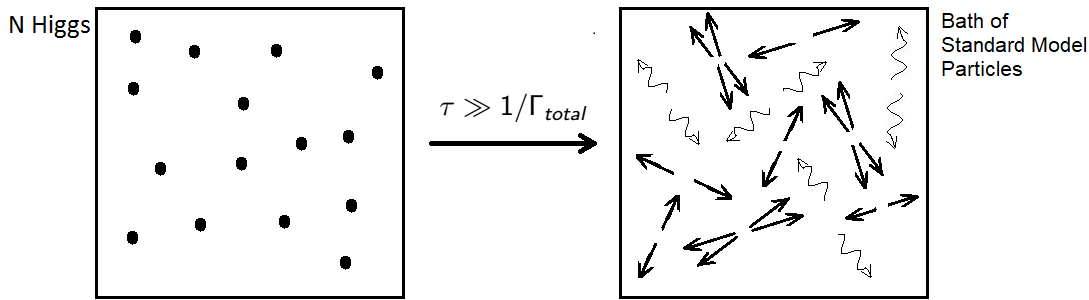}
  \caption{System of $N$ Higgs boson decaying into a bath of Standard Model particles.}
  \label{fig1}
\end{figure}

Consider an initial state of $N$ Higgs bosons that evolves, through  decay, 
to a final state composed by many different particles as represented in 
Fig.~\ref{fig1}. After a time 
$\tau\gg 1/\Gamma_{total}$ the system transforms into a thermal bath of particles from 
the Higgs boson decays, according to its branching ratios  
$BR_k(M_H|\pmb{\theta})={\Gamma_k(M_H|\pmb{\theta})}/{\Gamma_{total}(M_H|\pmb{\theta})}$, 
where $k=\gamma\gamma,\; gg,\; Z\gamma,\; q\bar{q},\; \ell^+\ell^-$, $ WW^*,\; ZZ^*$, 
are the 14 main SM decay modes; and $\pmb{\theta}$ represents all the other 
predetermined parameters entering into the formula; $\Gamma_k(M_H|\pmb{\theta})$ and 
$\Gamma_{total}(M_H|\pmb{\theta})$ are the partial 
and the total Higgs boson widths, respectively. The probability that 
the $N$ Higgs bosons system evolve to a given final state of a bath of Standard Model 
particles, defined by a partition of 
$n_1$ particles decaying into the mode $1$, $n_2$ particles decaying into the mode 
$2$, and so on until $n_m$ particles decaying into the mode $m$,  is given by 
\begin{equation}
\hspace{0.3 cm} P(\{n_k\}_{k=1}^m)\equiv \frac{N!}{n_1!\cdots n_m!}\prod_{k=1}^{m} 
\left[BR_k(M_H|\pmb{\theta})\right]^{n_k},  
\end{equation}
in which $\sum_i n_i=N$. 
We define the Shannon entropy \cite{shannon} associated to the decay of the 
system of $N$ Higgs bosons summing over each possible partition as 
\begin{equation}
S_N = -\sum_{\{n\}}^N P(\{n_k\}_{k=1}^m)\; \ln\left[P(\{n_k\}_{k=1}^m)\right]
= -\langle \ln(P)\rangle .
\label{a2}
\end{equation}

It can be shown, taking the asymptotic limit $N\rightarrow \infty$ formula for $S_N$~\cite{Cichon}, that maximization of $S_N$ is equivalent to the maximization of 
\begin{eqnarray}
S_\infty(M_H|\pmb{\theta})= \ln\left(\prod_{k=1}^m BR_k(M_H|\pmb{\theta})\right)
\label{Sinf}
\end{eqnarray}
The expression in Eq.~(\ref{Sinf}) is in fact the 
log-likelihood from the product of the branching ratios. 
Thus, we see as an outcome of MEP that the simultaneous maximization of the Higgs 
boson decays into 
Standard Model particles is equivalent to the maximization of the log-likelihood. 
\begin{figure}
\center \includegraphics[scale=0.37]{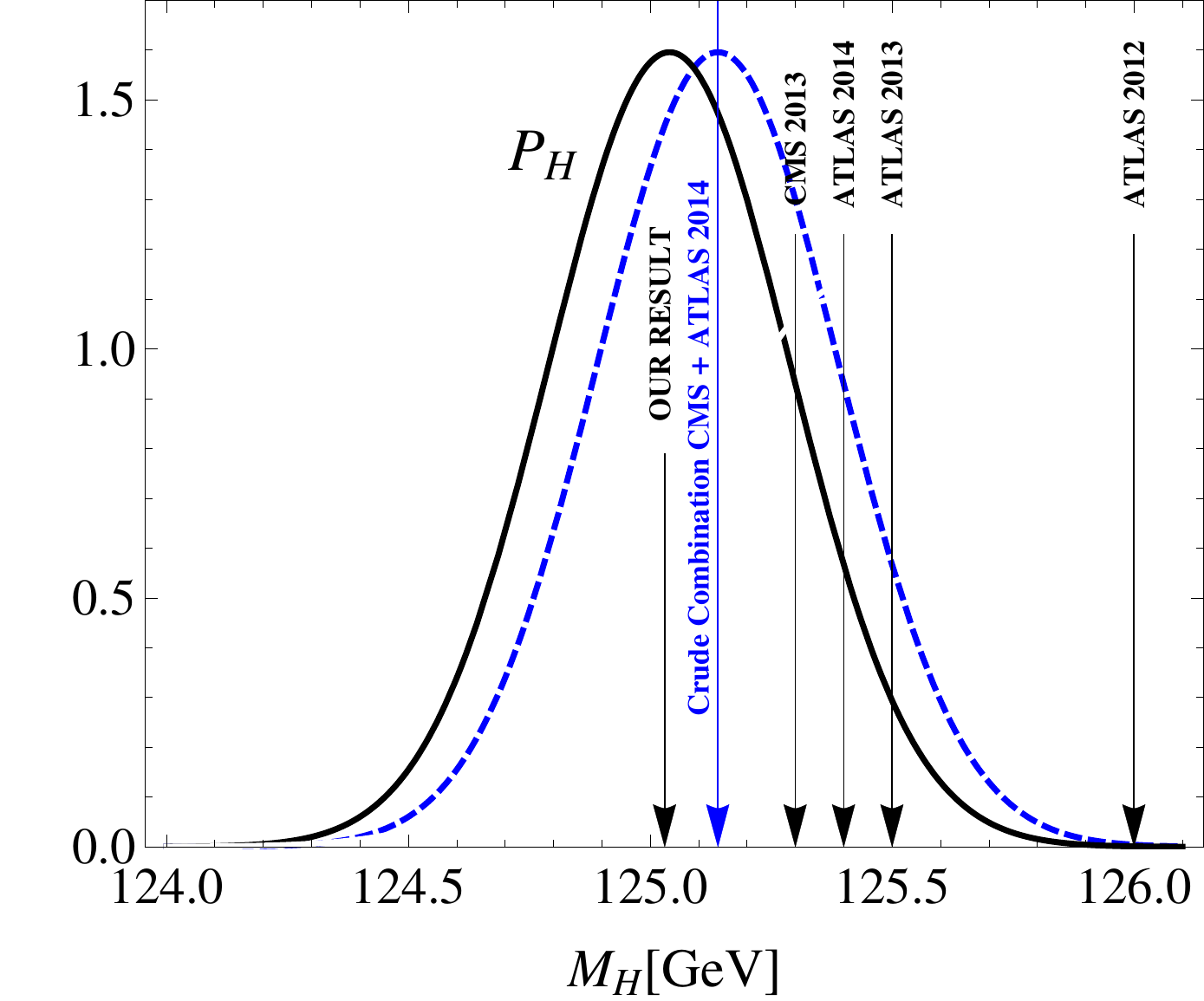}
\caption{Probability distribution function $P(M_H)$ obtained in 
\cite{Alves:2014ksa} from the maximum solution of $S_\infty$ is shown 
in the black curve. 
}
\label{fig2}
\end{figure}

Taking into account several sets of values within the experimental errors 
in the Standard Model electroweak parameters $\pmb{\theta}$~\cite{Dawson:2013bba}, 
a gaussian probability distribution function, $P_H(\hat{M}_H)$, is then fitted with 
the maximum solutions, $\hat{M}_H$, of $S_\infty(M_H|\pmb{\theta})$. The 
result is shown in Fig.~\ref{fig2}. The peak of $P_H(\hat{M}_H)$ is interpreted 
as the most probable value for the mass of the SM Higgs boson 
derived from MEP. Such a peak occurs for $M_H^{th} = 125.04\pm 0.25 \; \hbox{GeV}$, 
and reflects the combined uncertainties over the SM parameters.
It must be pointed out that this mass value determined from MEP is almost 
identical to the latest experimental value 
$M_H^{exp}= 125.09\pm 0.21\, (stat.)\,\pm 0.11\, (syst.)\,{\rm GeV}$
~\cite{Aad:2015zhl}. 

\subsection*{Inferring the axion mass}

We now review an application of MEP for inferring the mass of the hypothetical 
axion under two general 
assumptions. First, the axion can decay into pairs of neutrinos, in addition 
to the typical two photons decay channel. Second, the  effective 
Lagrangian describing the 
axion field, $A(x)$, interactions with the electromagnetic strength tensor, 
$F^{\mu\nu}$, and neutrinos fields, $\nu_{i}$, is 
\begin{equation}
\mathscr{L}_{eff}=\frac{1}{2}\partial _{\mu }A\,\partial ^{\mu }A-\frac{1}{2}%
m_{A}^{2}\,A^{2}-\frac{g_{A\gamma }}{4}\,A\,F^{\mu \nu }{\widetilde{F}}_{\mu
\nu }-\frac{g_{A\nu }}{2}\,\overline{\nu }_{i}\gamma ^{\mu }\gamma _{5}\nu
_{i}\,\partial _{\mu }A \,,
\label{effL}
\end{equation}
where $i=1,\,2,\,3$, and ${\widetilde{F}}_{\mu\nu }=\frac{1}{2}\epsilon_{\mu\nu\sigma\rho}F^{\sigma \rho }$. 
The axion-photon and the axion-neutrinos couplings are 
defined in terms of the axion decay constant, $f_{A}$, the fine structure 
constant, $\alpha$, and the order 1 model dependent anomaly coefficients, 
$C_{A\nu }$, ${\widetilde{C}}_{A\gamma }$, as 
$g_{A\gamma }={\alpha {\widetilde{C}}_{A\gamma }}/{2\pi f_{A}}$ and 
$g_{A\nu }={C_{A\nu }}/{f_{A}}$.
The effective Lagrangian in Eq.~(\ref{effL}) can be derived from a kind of 
DFSZ axion model with the addition of right-handed neutrinos~\cite{Alves:2017ljt}. 
It can be seen that the 
axion branching 
ratios $BR_{A\rightarrow\gamma\gamma}$ and $BR_{A\rightarrow\nu_i\nu_i}$ computed 
with Eq.~(\ref{effL}) depend only on the ratio $C_{A\nu }/{\widetilde{C}}_{A\gamma }$, 
the mass of the axion, $m_A$, and the lightest neutrino mass, $m_1$.

The neutrinos squared mass differences, 
$\Delta m_{12}^{2}=m_{2}^{2}-m_{1}^{2}=(7.45\pm 0.25)\times 10^{-5}\,\mathrm{eV}^{2}$ 
and 
$\Delta m_{31}^{2}=m_{3}^{2}-m_{1}^{2}=(2.55\pm 0.05)\times 10^{-3}\,\mathrm{eV}^{2}$~
\cite{King:2017guk}, 
assuming the normal hierarchy, enter as a prior information for the 
maximization of $S(m_{A}|m_{1 },r_{\nu })=\ln \left(BR_{0}BR_{1}BR_{2}BR_{3}\right)$. 
Still, there is the bound on the sum of neutrinos masses 
$0.059\text{ eV}<\sum_{i}m_{i}<0.23\;\text{eV}$, from the measurements 
of cosmic microwave background anisotropies~\cite{Ade:2015xua}, implying that 
 $0\leq m_{1 }<0.0712\;\text{eV}$. 

\begin{figure}
 \center \includegraphics[scale=0.29]{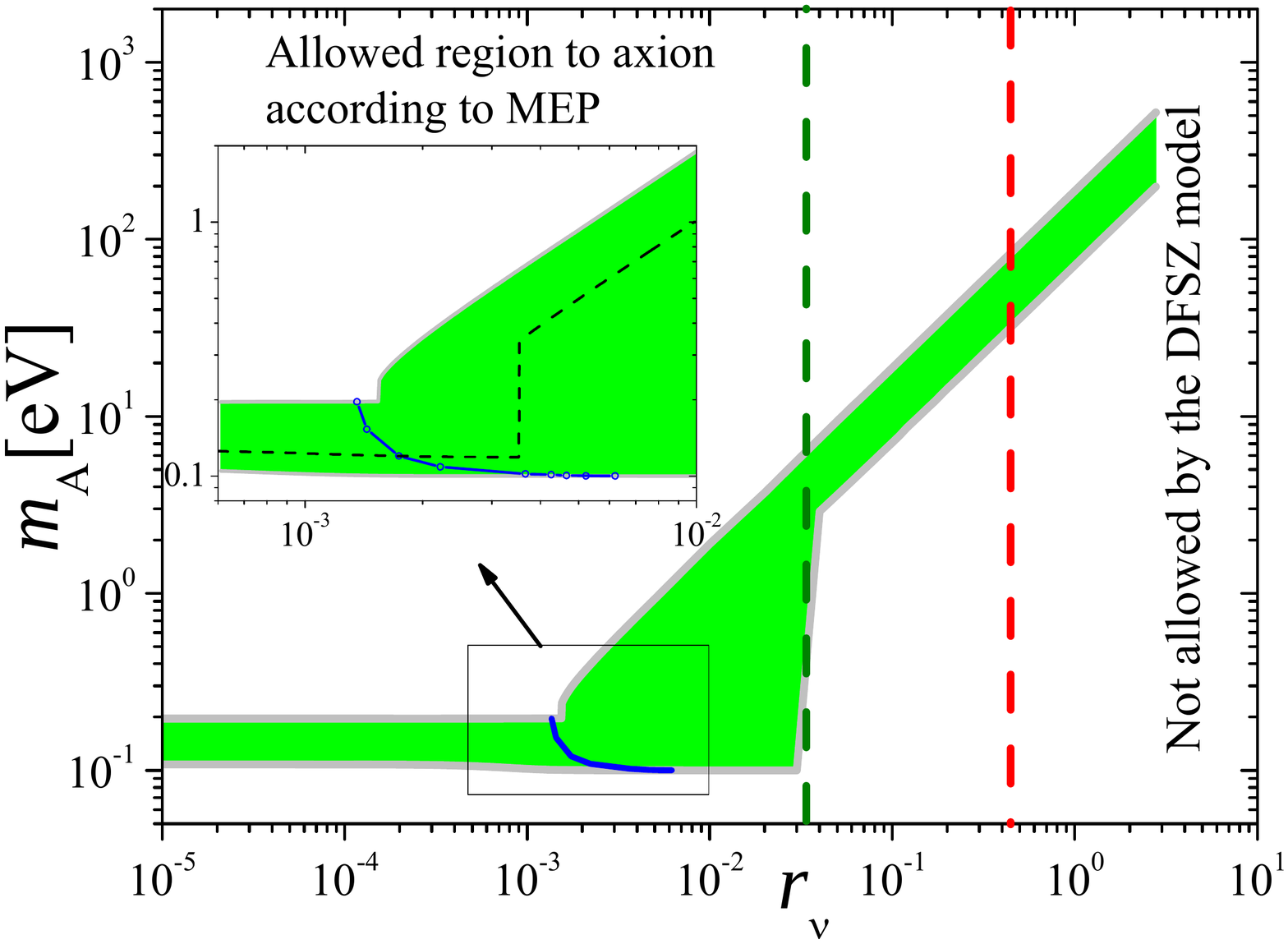}
\caption{Green region represents the maximum points of 
$S(m_{A}|m_{1 },r_{\nu })$ as obtained in 
Ref.~\cite{Alves:2017ljt}.    }
\label{fig3}
\end{figure}

It is shown in green on Fig.~\ref{fig3} the maximum points of the entropy 
$S(m_{A}|m_{1 },r_{\nu })$. The boundaries given by the gray curves 
are defined by the largest (upper) and lowest (lower) allowed value of  $m_1$. 
For the DFSZ model with right-handed neutrinos the anomaly coefficients 
are such that $r_\nu<0.46$ and the region beyond the red dashed line in 
Fig.~\ref{fig3} is excluded. Taking into account the astrophysical 
bounds from red giants~\cite{Viaux:2013lha} we obtain $r_\nu<0.034$, 
excluding the region beyond the green dashed line in Fig.~\ref{fig3}, 
and this implies that $0.1$~eV~$< m_{A}< 6.3$ eV. 

A more strong inference can be made maximizing the entropy as being a function 
of three unknowns, i. e., $S\equiv S(m_{A},m_{1 },r_{\nu })$. In this case 
we have a more sharp prediction of $0.1$~eV~$< m_{A}< 0.2$ eV, which is 
represented with the blue line in Fig.~\ref{fig3}. 

Finally, supposing that the axion can decay only into a pair of the 
lightest neutrino the inference made with MEP is a linear relation 
between $m_A$ and $m_1$, with the proportionality coefficient depending 
on $r_\nu$.

\vspace{0.25 cm}
{\bf{Acknowledgments}}: The authors thank CNPq and Fapesp. A. G. D. also thanks the organizers of Patras 2017 workshop, in special  Hero and Konstantin Zioutas for their hospitality.


\begin{footnotesize}

\end{footnotesize}


\end{document}